\documentclass{aastex702}

\usepackage{amsmath,amssymb,amsfonts}%
\usepackage{amsthm}%
\usepackage{mathrsfs}%
\usepackage{textcomp}%
\usepackage{booktabs}%

\begin{document}

\title{Wide Field Localization Using Fixed Modulation Collimators}

\author{Steven E. Boggs}
\affiliation{Department of Astronomy \& Astrophysics, University of California, San Diego, 9500 Gilman Drive, La Jolla, CA 92093, USA}
\affiliation{Space Sciences Laboratory, University of California, Berkeley, 7 Gauss Way, Berkeley, CA 94720, USA}
\email[show]{seboggs@ucsd.edu}
\correspondingauthor{Steven E. Boggs}

\shorttitle{Wide Field Localization with Fixed Modulation Collimators}
\shortauthors{Boggs}

\begin{abstract}
Rapid, precise localization of gamma-ray bursts (GRBs) requires continuous wide-field imaging at hard X-ray and soft gamma-ray energies. We develop
a fixed (non-rotating) modulation-collimator architecture to address this challenge. A bigrid collimator, a pair of aligned, periodic slit grids in front of a
simple, non-imaging photon-counting detector, transmits an off-axis source through a triangular fringe pattern whose local angular position, the source
phase, is recovered from simultaneous count rates in several grid pairs at fixed relative phase offsets, rather than by mechanically rotating a single
grid pair as in the classical rotating modulation collimator. Four-phase demodulation of these count rates robustly cancels detector background and grid leakage
without requiring rotation. Because the fringe pattern repeats across the sky, a phase measurement corresponds to many candidate source positions;
we resolve this ambiguity over an arbitrarily wide field of view using a cascade of grid stages with progressively longer beat periods (and pitches that converge toward a common finest value), generalizing the two-grid
vernier-scale technique originally used to localize Sco X-1. This vernier cascade technique reduces the required photon counts and grid-calibration accuracy
required to achieve fine angular resolution over a broad field of view. Generalized to two dimensions with orthogonal cascades, the resulting architecture achieves
wide-field localization with no moving parts, no time modulation, and only simple photon-counting detectors, a promising candidate for simple, inexpensive,
continuous-monitoring instruments supporting time-domain and multi-messenger astrophysics.
\end{abstract}


\section{Introduction}\label{sec1}

Rapid localization of gamma-ray bursts (GRBs) remains one of the central instrumental challenges of high-energy transient astronomy. Unlike a pointed
observation of a catalogued source, a burst arrives at an unpredicted time from a random direction, so the instrument must maintain sensitivity to accurate
positions over a wide field of view (FOV) rather than a narrow, pre-selected one. Three broad strategies have historically been used to meet this challenge. Coded-aperture
mask imagers (e.g., the Burst Alert Telescope on \emph{Swift} \citep{Gehrels2004}, the wide-field X-ray monitor on HETE-2 \citep{Kawai2003}, and, more recently, ECLAIRs on SVOM \citep{Godet2014,Wei2016}) achieve wide-FOV
imaging by replacing a single pinhole with a pseudo-random pattern of open and closed elements, achieving modest ($\sim$10--20\ifmmode'\else$'$\fi) angular
resolution ultimately limited by the resolution of the required position-sensitive detectors.
Multi-satellite time-of-flight triangulation networks localize bursts by comparing
arrival times across widely separated, non-imaging detectors \citep{Hurley2013, Burns2023}, which can typically achieve arcminute localizations, but typically with delays on order of a day.
The third strategy, the one this paper builds on, is the modulation collimator:
a pair of aligned, periodic slit grids in front of a non-imaging detector, whose transmission encodes a source's angular position through a fine, periodic fringe pattern
rather than through spatial imaging \citep{Oda1965}.

Historically (Sec.~\ref{sec2}), the modulation-collimator technique has been developed almost exclusively as a \emph{rotating} modulation collimator (RMC): a
single bigrid pair is spun about the detector boresight, and the resulting time-domain modulation is decoded, source by source, much as in Fourier or radio-interferometric
synthesis imaging. RMCs have delivered some of the finest angular resolutions ever achieved in hard X-ray/$\gamma$-ray astronomy \citep{Hurford2002}, but rotation
imposes either a rotating spacecraft or the mechanical complexity of rotating grid, and
time modulation is not ideally suited for rapidly varying sources such as GRBs, or short GRBs. The one clear precedent for using modulation collimators specifically for \emph{wide-field} burst
localization is the WATCH instrument, which was an RMC-based imager that relied on rotation of the entire instrument (grids and detectors) on its satellite platforms \citep{Lund1986, Lund1995}.

This paper develops an alternative approach to modulation collimators: a \emph{fixed} (non-rotating) bigrid modulation collimator architecture intended specifically for
wide-field GRB localization. In place of rotation, we use several bigrid pairs at fixed relative phase offsets to demodulate the source phase directly from simultaneous
count rates (Sec.~\ref{sec5}), and we resolve the resulting fringe ambiguity, the same many-fold ambiguity that motivated the original two-grid ``vernier''
technique of \citet{Gursky1966}, with a cascade of progressively coarser-pitched grid stages (Sec.~\ref{sec6}). The result is an architecture with no moving parts,
suited to 3-axis-stabilized platforms, that nonetheless achieves both a wide field of view and single-fringe source localization. Section~\ref{sec3} derives the geometry
and performance parameters of a single bigrid pair; Sec.~\ref{sec4} derives its transmission function; Sec.~\ref{sec5} develops single-, two-, and four-phase demodulation
of the source phase; Sec.~\ref{sec6} develops the multistage vernier cascade that turns a phase measurement into an unambiguous sky position across the full field of view;
and Sec.~\ref{sec7} generalizes this one-dimensional treatment to full two-dimensional sky localization and discusses the resulting architecture's advantages, including
its suitability for time-domain and multi-messenger astrophysics.

\section{Historical Context}\label{sec2}

The modulation collimator was introduced by \citet{Oda1965}, who replaced the single narrow slit of a conventional X-ray collimator with a pair of aligned, wide-open wire grids
of matched pitch. Because the two grids are far more open than a narrow-slit collimator of equivalent angular resolution, the detector's total counting rate, and hence its sensitivity,
becomes essentially independent of the angular resolution obtained, decoupling a trade-off that had previously constrained narrow-slit collimators. Oda's own stated motivation was
efficient observation of a source whose direction was already known approximately, not the localization of an unknown transient. The modulation collimator's ``broad field of view''
was from the outset an efficiency and pointing-tolerance advantage rather than an all-sky search capability.

The technique was extended almost immediately to precision source \emph{localization}. \citet{Bradt1968} generalized Oda's two-grid design to multigrid (three- and four-grid)
collimators that suppress redundant transmission peaks, and, together with \citet{Gursky1966}, flew a sounding-rocket payload carrying two four-grid collimators of slightly
mismatched period (a $\sim$5\% difference in band spacing) to measure the position of Sco~X-1 to $\sim$1$'$. Gursky et al.\ describe the underlying idea explicitly: ``the two
collimators provide a vernier scale,'' with the relative phase of transmission peaks between the two collimators identifying which particular fringe, among many nominally identical
ones, contains the source. This two-grid vernier technique is the direct conceptual ancestor of the multistage cascade developed in Sec.~\ref{sec6} of this paper, generalized
here from two stages to $N$.

Despite this introduction of the vernier technique, most of the field pursued time, rather than spatial, encoding. Building on a rotational encoding scheme proposed for optical imaging,
\citet{Schnopper1968} developed the rotating modulation collimator (RMC), a single bigrid pair is spun about the detector's boresight so that each source in the field produces a
distinct, geometry-dependent modulation frequency and phase that can be recovered by Fourier analysis of the time-series count rate, resolving multiple simultaneous sources
without requiring precision absolute pointing. \citet{Makishima1978} normalized the RMC technique and additionally introduced the multi-pitch modulation collimator (MPMC),
which combines several non-matched grid pitches and synthesizes an image via an inverse Fourier transform, explicitly analogous to aperture synthesis in radio astronomy.
This paper noted RMC imaging as ``especially suited to observing burst phenomena in a wide celestial field.''

That promise was demonstrated observationally by \citet{Nishimura1978}, who flew a balloon-borne rotating \emph{cross}-modulation collimator
(two identical bigrid pairs oriented $90^\circ$ apart, spun at 2~rpm) and obtained the first localization of a GRB by a single instrument (as opposed to inter-satellite
time-of-flight triangulation) to a precision of $\sim$0.3$^\circ$, from a $\sim$100$^\circ$$\times$100$^\circ$ field of view. \citet{Dean1983} subsequently reviewed
modulation collimators, RMCs, and coded-aperture masks on a common multiplex-advantage footing, cataloguing the achieved and required angular resolutions
across techniques, though by this point the field's wide-field/burst applications and its precision-imaging applications had already begun to diverge into largely
separate instrument lineages.

The wide-field lineage reached its fullest realization in WATCH \citep{Lund1995}, a spinning-instrument RMC monitor with a field of view exceeding $130^\circ$,
flown on the GRANAT and EURECA missions. Over its operational life WATCH identified more than 70 GRBs and localized roughly 40 of them to better than $1^\circ$,
and its authors explicitly frame the RMC's original role in precision source localization as having been ``taken over by the grazing incidence telescope systems,'' leaving wide-field
monitoring as the RMC's distinctive remaining niche. The precision-imaging lineage, meanwhile, continues to be widely used for narrow-FOV solar X-ray and gamma-ray imaging,
including HXT on SOLAR-A \citep{Kosugi1991}, RHESSI \citep{Lin2002}, STIX on Solar Orbiter \citep{Krucker2020}, and HXI on ASO-S \citep{Su2022}. 
RHESSI, whose nine co-aligned RMC subcollimators, spanning grid
pitches from 34~\textmu m to 2.75~mm, achieves angular resolutions from arcseconds to arcminutes and produced the first sub-arcminute image at $\gamma$-ray line energies
\citep{Hurford2002,Smith2004}. RHESSI's field of view, however, is only $\sim$1$^\circ$, essentially the opposite operating point from WATCH. Missions subsequent to WATCH,
which have required a wide field of view for burst-location missions (e.g., HETE-2/WXM, \emph{Swift}/BAT) have used coded-aperture masks,
not modulation collimators.

A separate, non-imaging lineage of wide-field burst monitors is also relevant here,
instruments that localize a burst not from a collimated fringe or coded shadow pattern, but purely from the relative count rates
recorded by several uncollimated scintillators pointed in different directions. BATSE \citep{Fishman1989} pioneered this approach with eight NaI detectors
at the corners of the \emph{Compton Gamma-Ray Observatory}, achieving typical burst localizations of several degrees (occasionally somewhat better for the
brightest events) from relative count-rate fitting alone. Its successor, Fermi-GBM \citep{Meegan2009}, extends the same technique with improved sensitivity, and the
CubeSat-class BurstCube mission \citep{Racusin2017} demonstrates that this relative-count-rate approach can be implemented on a low-cost small-satellite platform.
Because these instruments share our reliance on simple, non-imaging photon-counting detectors rather than a spatially resolved image, and because they
represent the low-cost end of the wide-field burst-monitor design space, we return to a direct comparison with them in Sec.~\ref{subsec7-2}.

In terms of modulation collimators, the literature reflects, on one side, a mature and precise but narrow-field RMC technique, and on the other, a single dedicated wide-field RMC burst
monitor (WATCH) that existed
before the era of rapid localizations. Comparatively little attention has been paid to reviving the original, purely geometric vernier disambiguation of \citet{Gursky1966},
which requires no rotation at all, as the basis for a wide-field instrument. That is the gap this paper addresses. Sections~\ref{sec3}-\ref{sec6} develop a fixed, multi-stage
bigrid architecture that extends the two-grid vernier concept to an arbitrary number of cascaded stages, aiming to combine RMC-like source-position precision
with WATCH-like field of view, but requiring no moving parts or platform rotation and utilizing simple photon-counting detectors.

\begin{figure}
\centering
\plotone{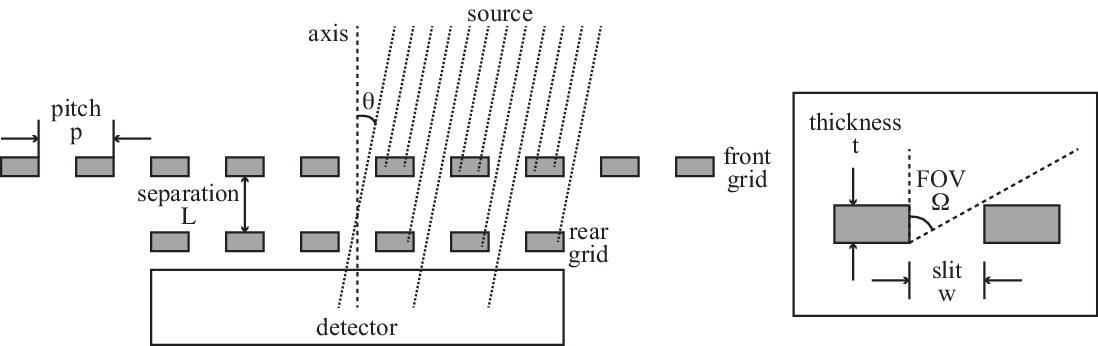}
\caption{\label{fig:f1} Representative geometry of a bigrid pair mounted in front of a photon-counting detector, showing characteristic sizes.
The box at the right shows a blow up of a single grid, demonstrating the self-collimating FOV of the grid. Adapted from \citep{Hurford2002}.}
\end{figure}

\section{Principle of the Static Bigrid Modulation Collimator}\label{sec3}

\subsection{Geometry}\label{subsec3-1}

A bigrid collimator consists of a pair of 1D parallel-slit grids, separated by a distance $L$, mounted in front of a non-imaging, photon counting detector (Fig.\,\ref{fig:f1}). The grids, of thickness $t$, are opaque to the incident (X-ray or gamma-ray) radiation and are composed of parallel slits of pitch $p$ and slit width $w$, with $w = p/2$ in the standard bigrid design. In what follows we assume the slits of the upper and lower grids are perfectly aligned; we will discuss introducing a relative shift between the grids below (Sec.~\ref{sec5}). The relevant physical parameters of the grids are the \emph{grid pitch} $p$, \emph{slit width} $w$, \emph{grid separation} $L$, and \emph{grid thickness} $t$.

From these four parameters we can define three performance parameters for the grid pair:
\begin{align}
\Delta\theta &= \tan^{-1}\!\left(\frac{w}{L}\right) \approx \frac{w}{L},
  \label{eq:dtheta}\\[4pt]
\alpha &= \tan^{-1}\!\left(\frac{p}{L}\right) \approx \frac{p}{L},
  \label{eq:alpha}\\[4pt]
\Omega &= \tan^{-1}\!\left(\frac{w}{t}\right).
  \label{eq:omega}
\end{align}%

Equation~\eqref{eq:dtheta} is the classic definition of the \emph{angular resolution} $\Delta\theta$ of bigrid collimator (Sec.~\ref{sec4}); Eq.~\eqref{eq:alpha} is the angular \emph{modulation period} $\alpha$ 
(or fringe period) with which the transmission pattern repeats across the sky; and Eq.~\eqref{eq:omega} is the self-vignetting angle $\Omega$ at which the finite grid thickness occludes off-axis rays passing through a single grid, independent of the second grid. As used in Sec.~\ref{sec6}, $\Omega$ also sets the half-width of the usable FOV over which fringes must be counted and disambiguated.

\section{Transmission Through a Bigrid Pair}\label{sec4}

\subsection{Ideal (Opaque) Grid Transmission}\label{subsec4-1}

\begin{figure}
\centering
\plotone{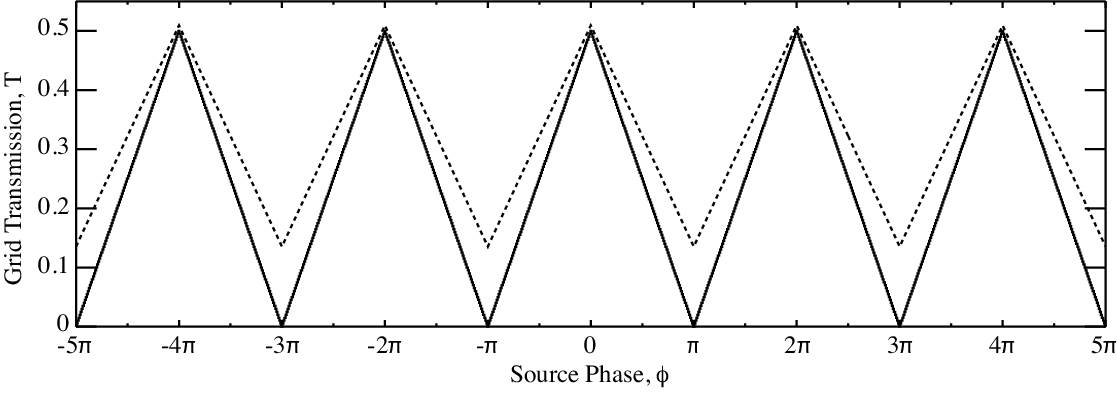}
\caption{\label{fig:f2} The ideal transmission function for a bigrid pair (solid line), compared to a realistic transmission function assuming $\mu x = 2$ (dashed line).
The realistic transmission function largely retains the sawtooth shape of the ideal function, but with reduced amplitude and a nearly-constant leakage baseline.}
\end{figure}

Consider a source at an off-axis angle $\theta$ (\emph{source angle}), measured in the one-dimensional offset direction perpendicular to the grid slits. For ideal (perfectly opaque) grids, the transmission $T$ through the bigrid pair is a triangular (sawtooth) function of $\theta$ with angular period $\alpha$, peaking at $T = 1/2$ when the slits of the two grids are aligned ($\theta = 0, \pm\alpha, \pm 2\alpha, \ldots$) and
falling to $T = 0$ at angles where the slats of the upper grid are aligned with the slits of the lower grid ($\theta = \pm\tfrac{\alpha}{2}, \pm\tfrac{3\alpha}{2}, \ldots$), as demonstrated in Figure \ref{fig:f2}.

It is convenient to define the \emph{source fringe offset} $\delta$, the signed angular displacement of the source from the center of the nearest fringe,
\begin{align}
\delta &\equiv \left[\left(\theta + \frac{\alpha}{2}\right) \bmod \alpha\right] - \frac{\alpha}{2},
\qquad -\frac{\alpha}{2} < \delta \le \frac{\alpha}{2},
\label{eq:offset}
\end{align}%
and the corresponding normalized \emph{source phase} $\phi$,
\begin{align}
\phi &\equiv \frac{2\pi\delta}{\alpha}, \qquad -\pi < \phi \le \pi,
\label{eq:phase}
\end{align}%
which is the quantity actually measured by a single bigrid pair. Unlike $\delta$, which retains the angular units of $\theta$ and $\alpha$ and whose numerical range depends on the grid pitch, $\phi$
is dimensionless and spans the same fixed interval $(-\pi,\pi]$ for every grid pair regardless of pitch, making it directly comparable across the different modulation periods used in the
vernier cascade (Sec.~\ref{sec6}).
Using this definition of $\phi$, the ideal transmission function is given by the simple equation,
\begin{align}
T(\phi) = \tfrac{1}{2} \left[1-\frac{|\phi|}{\pi}\right].
\label{eq:trans-ideal}
\end{align}%
Because the transmission pattern is periodic in $\theta$ with period $\alpha$, a measurement of $\phi$ (equivalently, of $\delta$) corresponds to a set of candidate source positions, the \emph{fringes} $\eta_k$,
\begin{align}
\eta_k &= k\alpha + \delta, \qquad k = 0, \pm 1, \pm 2, \ldots,
\label{eq:fringes-single}
\end{align}%
where the true source position $\theta = \eta_k$ for exactly one value of $k$. Resolving which fringe corresponds to the true source position is the localization problem addressed in Sec.~\ref{sec6}.

We introduce the normalized, repeating sawtooth function $h(\phi)$, with $-1 \le h(\phi) \le 1$, defined by $h(0) = 1$, $h(\pm\pi) = -1$, $h(\phi + 2\pi) = h(\phi)$.
The ideal transmission function in terms of this sawtooth function is,
\begin{equation}
T(\phi) = \frac{a}{2}\!\left[h(\phi) + 1\right],
\label{eq:T-ideal}
\end{equation}%
where $a = \frac{1}{2}$ for an ideal (opaque) grid. A source of flux $\mathcal{F}$ at off-axis angle $\theta$ produces a count rate in the detector
\begin{equation}
R = T(\phi)\,\mathcal{F} A = \frac{a}{2}\,h(\phi)\,\mathcal{F} A + \frac{a}{2}\,\mathcal{F} A,
\label{eq:R-ideal}
\end{equation}%
where $A$ is the effective area of the photon collecting detector in the absence of the grids.

\subsection{Real (Imperfect) Grid Transmission}\label{subsec4-2}

A real grid is not perfectly opaque. Leakage through the grid is characterized by the factor $\mu x$, where
$\mu$ is the linear attenuation coefficient for the specific photon energy, and $x$ is an \emph{average} path length
of an incident photon through a single grid. The value for $x$ will vary across the FOV of the grids, but here we
consider a single average value for simplification. Using this factor, we can modify our transmission function to account
for leakage,
\begin{align}
T(\phi) = \tfrac{1}{2} \left[1-\frac{|\phi|}{\pi}\right](1+e^{- 2 \mu x}) + \frac{|\phi|}{\pi}e^{- \mu x},
\label{eq:trans-real}
\end{align}%
where $T(\phi)$ approaches the ideal transmission function [Eq.~\eqref{eq:trans-ideal}] in the limit $\mu x \gg 1$.

We can rewrite this transmission function using the same sawtooth formalism used above,
\begin{equation}
T(\phi) = \frac{a}{2}\!\left[h(\phi) + 1\right] + \ell,
\label{eq:T-real}
\end{equation}%
where $h(\phi)$ is the same normalized sawtooth as above ($-1 \le h(\phi) \le 1$), $a \le 1/2$ is the reduced modulation amplitude due to partial grid transparency, and $\ell$ is a leakage term:
\begin{align}
a &= \frac{1}{2} \left( 1-e^{-\mu x}\right)^2,
  \label{eq:leak-a}\\[4pt]
\ell &= e^{-\mu x}.
  \label{eq:leak-ell}
\end{align}%

This leakage term, while generally assumed to be small ($\ell \ll 1$), is not necessarily trivial. This term adds another level of systematic uncertainty to identifying $\phi$ and $\theta$,
which can be avoided using a multiple-phase bigrid design (Sec.~\ref{subsec5-3}).

Including a background count rate $\mathcal{B}$, the measured detector rate is,
\begin{equation}
R = T(\phi)\,\mathcal{F} A + \mathcal{B} = \frac{a}{2}\,h(\phi)\,\mathcal{F} A + \left[\frac{a}{2} + \ell\right] \mathcal{F} A + \mathcal{B}.
\label{eq:R-measured}
\end{equation}%

To summarize, for a single bigrid pair of modulation period $\alpha$, the measured count rate in the detector depends on the source phase $\phi$.
Measuring $R$ measures $\phi$, and hence the fringe offset $\delta = \alpha\phi/2\pi$; the resulting fringes repeat with period $\alpha$ across the sky [Eq.~\eqref{eq:fringes-single}].
The true source position lies at one of these fringes.

\subsection{Field-of-View Constraints}\label{subsec4-3}

The transmission function derived above describes the fine (fringe-scale) response of the grid pair. 
For wide field applications, the transmission function must account for \emph{self-vignetting} of the grids \citep{Gaither1996, SaintHilaire2024}, 
i.e., a grid slat thick enough to be opaque at hard X-ray and gamma-ray energies inevitably clips off-axis photons. 
Self-vignetting has two primary effects. First, this effect will ultimately constrain the FOV in the direction perpendicular to the slits. 
This is exemplified in Eq.~\eqref{eq:omega} for rectangular grid slats. Second, self-vignetting can flatten the transmission function,
potentially leading to blind spots in the FOV. These effects can be largely reduced and potentially eliminated altogether with 
careful grid design, which we investigate in detail in a companion paper \citep{Boggs2026}. Hence we largely ignore the 
effects of self-vignetting in this work.

A related, purely geometric effect is the lateral shift of the front grid's shadow across
the rear grid (and detector) with source angle, which grows as $L\tan\theta$. Unless the front grid is made correspondingly
larger than rear grid, following the precedent of oversizing a coded mask relative to its detector plane, the FOV over which the two grids'
patterns fully overlap will be limited by the solid angle of the front grid relative to the rear grid. In this paper we
assume the upper grid is large enough to not constrain the FOV.

Finally, because a single 1D grid pair provides self-vignetting only in the direction perpendicular to its own slits, it provides no
thickness vignetting at all along the slit direction itself. Each cascade (an $X$- or $Y$-oriented system, Sec.~\ref{subsec7-1}) is therefore, in principle,
exposed to sky background from well beyond $\pm\Omega$ along its own slit direction unless additional collimation or shielding, orthogonal to the vernier
grids themselves, is introduced to bound this direction as well. We do not assume any such collimation in this work.

\section{Demodulation: Measuring the Source Phase}\label{sec5}

Given the angular dependence of the measured count rate derived above, we can explore how best to recover the source phase $\phi$ from a measured count rate.

\subsection{Single-Phase Demodulation}\label{subsec5-1}

With a single bigrid collimator, the measured rate in the detector is
\begin{equation}
R_1 = \frac{a}{2}\,h(\phi)\,\mathcal{F} A + \left[\frac{a}{2} + \ell\right] \mathcal{F} A + \mathcal{B}_1,
\label{eq:R1-single}
\end{equation}%
where the subscript ``1'' anticipates the addition of further bigrid pairs below. To recover $\phi$ from $R_1$ alone requires independent knowledge of the background rate $\mathcal{B}_1$, the transmission and leakage properties ($a$, $\ell$), and the source flux $\mathcal{F}$.

Requiring independent knowledge of $\mathcal{F}$ is undesirable; in practice $\mathcal{F}A$ can be measured with a second, unshielded (uncollimated) monitor detector. Provided $\mathcal{B}_1$ is known or can be modeled,
and $\ell$ is either negligible or well characterized, these contributions to the overall rate can be subtracted, giving the residual rate $R_1'$,
\begin{equation}%
R_1' = \frac{a}{2}\,h(\phi)\,\mathcal{F} A.
\label{eq:single-phase-residual}
\end{equation}%
A subtlety with this single-phase measurement is that the triangular transmission function is symmetric about each peak, so $h(\phi) = h(-\phi)$, effectively doubling the number of fringes that are created, and
potentially leading to confusion in deriving the source location. This degeneracy is easily broken by the introduction of a second bigrid/detector pair.

\subsection{Two-Phase Demodulation}\label{subsec5-2}

The measurement of $\phi$ becomes more robust if a second bigrid/detector pair is introduced, with the \emph{same} modulation period $\alpha$ but with its grids shifted relative
to each other in the direction perpendicular to the slits,
producing a phase difference $\Pi$, expressed in the same normalized (radian) units as the source phase $\phi$ [Eq.~\eqref{eq:phase}], in the transmission curve. For example, shifting
one grid by $p/4$, a quarter of the physical grid pitch, introduces a phase shift $\Pi = \pi/2 = 90^\circ$, i.e., one quarter of the full $2\pi$ modulation cycle regardless of the pitch $p$ itself.

Writing the measured rate for each bigrid pair as
\begin{equation}
R_i = T(\phi - \Pi_i)\,A \mathcal{F} + \mathcal{B}_i,
\label{eq:Ri-general}
\end{equation}%
where $\Pi_i$ is the grid phase shift of grid pair $i$, most designs adopt $\Pi_1 = 0^\circ$ and $\Pi_2 = 90^\circ$, giving a sine/cosine-like pair of measured rates:
\begin{align}
R_1 &= \frac{a}{2}\,h(\phi)\,\mathcal{F} A + \left[\frac{a}{2} + \ell\right] \mathcal{F} A + \mathcal{B}_1, \\
R_2 &= \frac{a}{2}\,h(\phi - \Pi_2)\,\mathcal{F} A + \left[\frac{a}{2} + \ell\right] \mathcal{F} A + \mathcal{B}_2.
\end{align}%
Here we have assumed that the grid response properties ($a$, $\ell$) are the same for both sets of grids. This is not a requirement, and these properties could be tracked in
these equations for each brigrid pair ($a_1$, $\ell_1$, etc.), with differences calibrated out at the end. However, microscopic variations in grid manufacturing will tend to average out over the
full grid, making such variations small compared to our required accuracy (Sec.~\ref{sec6}).

Two-phase demodulation relies on $\mathcal{B}_1$ and $\mathcal{B}_2$ being characterized and subtractable, and on $\ell$ being either negligible or well characterized. If these terms are subtracted, the remaining residual rates $R_1'$, $R_2'$ yield,
\begin{equation}
\frac{R_1'}{R_2'} = \frac{h(\phi)}{h(\phi - \Pi_2)}.
\label{eq:two-phase-ratio}
\end{equation}%
The ratio of the residual rates thus yields a unique measurement of $\phi$.

Two-phase measurements no longer require independent knowledge of $\mathcal{F}$, and avoids the $h(\phi) = h(-\phi)$ symmetry noted for the single-phase measurement.
Even the dependence on the modulation amplitude, $a$, technically drops out of this measurement.
However, deriving the residual rates $R_i'$ required for two-phase measurements
still requires knowing or estimating the background rates $\mathcal{B}_i$,
and that $\ell$ is either negligible or well characterized.

\begin{figure}
\centering
\plotone{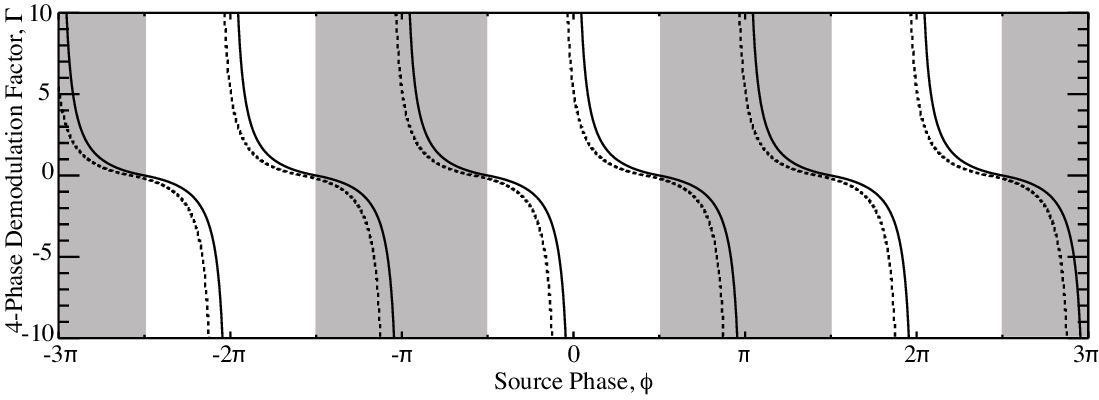}
\caption{\label{fig:f3} Four-phase demodulation factor $\Gamma$ [Eq.~\eqref{eq:four-phase-ratio}] for the ideal grids (solid lines) and realistic grids ($\mu x = 2$, dashed lines).
The realistic grid response is shifted to separate the curves, which are otherwise indistinguishable. The factor as defined repeats
with a phase frequency of $\pi$, but is uniquely distinguishable over the full $2\pi$ source phase by distinguishing where $R_1 -R_3 > 0$ (white bands)
and $R_1 -R_3 < 0$ (gray bands).}
\end{figure}

\subsection{Four-Phase Demodulation}\label{subsec5-3}

The measurement of $\phi$ becomes more robust still if four bigrid/detector pairs are used, each with the same modulation period $\alpha$ but with different phase offsets $\Pi_i$.
The power of this method follows from the symmetry of the grid response,
\begin{align}
&h(\phi - 180^\circ) = -h(\phi), \\
&\ell = \text{const.}
\end{align}%

Take the particular case
$\Pi_1 = 0^\circ$, $\Pi_2 = 90^\circ$, $\Pi_3 = 180^\circ$, $\Pi_4 = 270^\circ$. Consider the difference in rate between opposite pairs:
\begin{align}
R_1 - R_3 = {}& \frac{a}{2}\left[h(\phi) - h(\phi - 180^\circ)\right] \mathcal{F} A \nonumber\\
             & + \left[\ell - \ell\right] \mathcal{F} A + (\mathcal{B}_1 - \mathcal{B}_3).
\end{align}%
This rate difference enables significant simplification in determining the source phase $\phi$, within statistical uncertainties:
\begin{align}
&\text{(i) } \mathcal{B}_1 - \mathcal{B}_3 = 0, \\
&\text{(ii) } \ell - \ell = 0 \quad (\ell \text{ is the same constant in every channel}), \\
&\text{(iii) } h(\phi) - h(\phi - 180^\circ) = 2\,h(\phi).
\end{align}%
The rate difference between two pairs with grids shifted by $180^\circ$ therefore allows the background \emph{and} the slit leakage to be directly subtracted:
\begin{equation}
R_1 - R_3 = a\,h(\phi)\,\mathcal{F} A.
\label{eq:R1minusR3}
\end{equation}%
Likewise,
\begin{equation}
R_2 - R_4 = a\,h(\phi - 90^\circ)\,\mathcal{F} A,
\label{eq:R2minusR4}
\end{equation}%
leaving our \emph{four-phase demodulation factor}, $\Gamma$:
\begin{equation}
\Gamma = \frac{R_1 - R_3}{R_2 - R_4} = \frac{h(\phi)}{h(\phi - 90^\circ)}.
\label{eq:four-phase-ratio}
\end{equation}%
Not only do four-phase measurements not require an independent measure of $\mathcal{F}$, they no longer require
knowing the background rates $\mathcal{B}_i$, nor depend on the leakage term $\ell$ being either negligible or well characterized.

Figure \ref{fig:f3} shows this four-phase demodulation factor $\Gamma$ [Eq.~\eqref{eq:four-phase-ratio}] for both the ideal transmission function and the realistic transmission function.
This figure demonstrates that four-phase demodulation is robust and insensitive to grid leakage for modest levels of leakage, providing a clean and unique measurement of the
source phase $\phi$.

The four-phase measurement also admits a clean, closed-form statistical
uncertainty on the recovered fringe position $\eta$. Under an idealized (background-free, ideal-transmission-model) treatment, the four channel counts are pure
Poisson measurements of the transmitted source flux, and standard error propagation through the four-phase ratio [Eq.~\eqref{eq:four-phase-ratio}] gives the uncertainty $\sigma_\eta$ on the
fringe angle $\eta$ (four-phase, ideal transmission, background-free), 
\begin{equation}
\sigma_\eta = \frac{\Delta\theta}{\sqrt{2S}},
\label{eq:sigma-eta-four}
\end{equation}%
where $S$ is the total number of source counts combined across all four channels of a single four-phase module.
Except for the exact numerical factor of $1/\sqrt{2}$, this dependence on $\Delta\theta$ and $S$ agrees with expectations
from typical photon-limited measurements. Equation~\eqref{eq:sigma-eta-four} is the origin of the $1/\sqrt{S}$ precision scaling used in Sec.~\ref{subsec6-1} to
relate the required source counts to the disambiguation factor $D$; the full derivation is given in Appendix~\ref{appA}.

\section{Localization and the Vernier Grid Cascade}\label{sec6}

Once the source phase $\phi_j$ has been measured from a single modulation period $\alpha_j$ [equivalently, the fringe offset $\delta_j = \alpha_j\phi_j/2\pi$, Eqs.~\eqref{eq:offset}--\eqref{eq:phase}],
the source \emph{position} has not yet been measured: what has been measured is a periodic series of fringes $\eta_j$ across the field of view (Fig. \ref{fig:f4}), 
exactly one of which corresponds to the true source position [cf. Eq.~\eqref{eq:fringes-single}],
\begin{equation}
\eta_{j,k} = k\alpha_j + \delta_j.
\label{eq:fringes-j}
\end{equation}%
Counting the number of fringes across a genuinely wide field of view requires more care than the small-angle forms of Sec.~\ref{subsec3-1} provide: each grid's transmission pattern is periodic not in the source angle $\theta$ itself, but in the projected shadow shift across the lower grid, which scales as $\tan\theta$ rather than $\theta$. The fringe count over the full field of view $[-\Omega,\Omega]$ must therefore be taken in $\tan\theta$-space. Given a full field of view $2\Omega$, the number of fringes $M_j$ (candidate positions) across the field of view at period $\alpha_j$ is
\begin{equation}
M_j = \frac{2\tan(\Omega)}{\tan(\alpha_j)}.
\label{eq:Mj}
\end{equation}%
For the grid pitches $\alpha_j$ themselves, which remain small (of order a degree or less) throughout this paper, $\tan\alpha_j\approx\alpha_j$ to high accuracy; it is only $\Omega$, and the beat periods $\beta_{1,j}$ introduced below, that grow large enough over the cascade to require this distinction.

To turn a phase measurement into an actual source location, the correct fringe must be identified. This is done using the \emph{vernier cascade} method
\citep{Gursky1966}.

\subsection{Fringe Ambiguity and the Disambiguation Factor}\label{subsec6-1}

\begin{figure}
\centering
\plotone{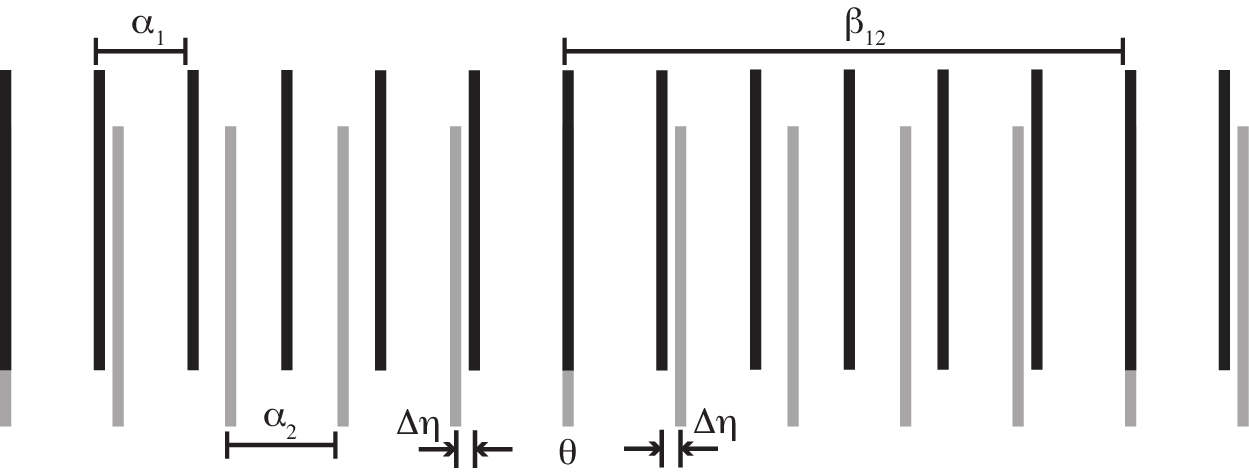}
\caption{\label{fig:f4} Schematic of a single-stage vernier cascade, where $\alpha_2 \approx 1.25 \alpha_1$, leading to a beat frequency $\beta_{12} \approx 5 \alpha_1$
and a per-stage disambiguation factor $d \approx 5$.
The primary fringes ($\alpha_1$) are shown in black, and the secondary fringes ($\alpha_2$) are shown in gray.
The overlapping fringes corresponding the true source location are labeled with $\theta$. The separation between the neighboring fringes, $\Delta \eta$, is also illustrated.
From the vernier cascade, every beat period a single fringe remains a potential candidate location,
while the other $d-1$ fringes within the beat period can be rejected with high confidence.}
\end{figure}

In the simplest form of a vernier cascade, a second set of grids of period $\alpha_2 \ge \alpha_1$ can be introduced, chosen to create a \emph{beat period} between the fringes of the two grids. Because the fringes of both grids are periodic in $\tan\theta$ rather than $\theta$ (as above), the beat period $\beta_{12}$ is set by the corresponding condition in $\tan\theta$-space,
\begin{equation}
\tan(\beta_{12}) = \frac{\tan\alpha_1\tan\alpha_2}{\tan\alpha_2 - \tan\alpha_1} \approx \frac{\alpha_1 \alpha_2}{\alpha_2 - \alpha_1},
\label{eq:beat-period}
\end{equation}%
where the last step uses $\tan\alpha_j\approx\alpha_j$ for the small grid pitches, exactly as in Eq.~\eqref{eq:Mj}; the beat period itself is then $\beta_{12}=\arctan[\cdots]$, which need not be small. The fringes of the two systems will align at the true source position, and will otherwise only align roughly every beat angle $\beta_{12}$ relative to the true position (Fig. \ref{fig:f4}).
If the design is chosen such that,
\begin{equation}
\tan(\beta_{12}) = 2\tan(\Omega),
\label{eq:beat-fov}
\end{equation}%
the beat pattern aligns the fringes only \emph{once} over the entire field of view, at the correct source position.
This is equivalent to requiring, $M_2 = M_1 - 1$,
i.e., the second grid system covers the entire field of view with one fewer fringe than the first system, which ensures that only one fringe aligns between the two systems at the correct source position.

The challenge lies in the statistics of how well the aligned fringe can be distinguished from the unaligned fringes. We define the \emph{disambiguation factor} $D$,
\begin{equation}
D \equiv M_1 = \frac{2 \tan(\Omega)}{\tan(\alpha_1)},
\label{eq:D}
\end{equation}%
which is the number of ambiguous fringes that must be sorted through to find the real one.
The larger $D$ is, the more statistical precision is required to differentiate the true (aligned) fringe from its $D-1$ neighbors.

This precision requirement can be made explicit using this single-stage beat pair. Because $\tan(\beta_{12})=2\tan(\Omega) \approx D\alpha_1$ [Eq.~\eqref{eq:beat-fov}, \eqref{eq:D}],
we can rearrange Eq.~\eqref{eq:beat-period}
to derive $\alpha_1/\alpha_2 = 1-1/D$: the two grid periods are nearly identical, differing fractionally by only $1/D$.
The fringes between the two grids will be aligned (i.e., $\Delta \eta = |\eta_2-\eta_1| = 0$) at the source location $\theta$. 
The separation between the next set of \emph{neighboring} fringes (Fig.\,\ref{fig:f4}) will be given by,
\begin{equation}
\Delta \eta = |\eta_2-\eta_1| = \alpha_2 - \alpha_1 \approx \frac{\alpha_1}{D}.
\label{eq:fringe_spacing}
\end{equation}%

In order to distinguish this neighboring fringe misalignment at the level of $k$ standard deviations, i.e., the $k\sigma$ level, we require the uncertainty on the fringe difference
measurement to meet the requirement:
\begin{equation}
k \sqrt{\sigma_{\eta 1}^2 + \sigma_{\eta 2}^2} \lesssim \Delta \eta \approx \frac{\alpha_1}{D}.
\label{eq:precision-required}
\end{equation}%
Taking the approximation $\sigma_{\eta 2} \approx \sigma_{\eta 1} = \alpha_1/(2\sqrt{2S})$ [Eq.~\eqref{eq:sigma-eta-four}], we can rearrange to find the minimum source count
to guarantee disambiguation of these fringes,
\begin{equation}
S \gtrsim \left( \frac{k}{2} \right)^2 D^2.
\label{eq:S-min-D}
\end{equation}%
The required value of $k$ depends on the confidence level ($CL$) at which we want to disambiguate these fringes against false alignments. 
In Appendix~\ref{appC}, we demonstrate that for $CL \sim 99\%$, we anticipate $k \sim 2.5-3.0$.
This is the scaling adopted in the worked example of Sec.~\ref{subsec6-4}.

Distinguishing the neighboring fringe misalignment also places a constraint on the accuracy of the grid response, i.e., the calibration accuracy.
Because this accuracy reflects a systematic, fixed uncertainty in the calibrated grid geometry, it does not improve with additional source counts $S$
the way the statistical precision of Eq.~\eqref{eq:S-min-D} does. For the calibration accuracy to not become the limiting factor in fringe disambiguation
regardless of exposure, it must satisfy the upper limit,
\begin{equation}
\epsilon \ll \frac{1}{D}
\label{eq:accuracy-one-stage}
\end{equation}%
This accuracy is a requirement on the calibrated relationship between the phase
measured in a given module and the sky angle it corresponds to, i.e., on how accurately the grid pitches $\alpha_j$ and the grid separation $L$ of every stage are
characterized \emph{relative} to one another. This requirement is in addition to, and independent of, the statistical precision of any single phase measurement (which is set by $S$, as above, and can always be improved with more exposure). We discuss this
requirement further in Sec.~\ref{subsec6-2}.
If $D \gg 1$, the accuracy (calibration) and precision (photon statistics) required to differentiate aligned and unaligned
fringes becomes very challenging.
For reference, typical calibration accuracies for hard X-ray and gamma-ray instruments are $\sim 1 - 2\%$.
Correspondingly, a rough rule of thumb is that $D \lesssim 10$ is readily feasible, while $D \gtrsim 100$ is challenging with a single-stage vernier,
both in terms of the required accuracy of the grid calibration
as well as the number of source counts required to meet the statistical precision.

Eq.~\eqref{eq:S-min-D} is derived in the source-dominated limit
$S \gg B$ [Appendix~\ref{appA}, Eq.~\eqref{eq:sigma-eta-four-app}]. The general, background-inclusive uncertainty [Eq.~\eqref{eq:sigma-eta-four-app-general}] should be used in place of
Eq.~\eqref{eq:sigma-eta-four} whenever $B$ is not small compared to $S$. Translating this requirement into a peak-flux threshold, or into
the fraction of a real GRB population (e.g., as tabulated by \emph{Swift}-BAT or Fermi-GBM) that would be localized to a given accuracy, requires
a detailed energy response, off-axis transmission and effective area, and a background estimate for a specific orbit and detector design, all of which are beyond
the scope of this introduction to the fixed modulation collimator technique. We will present 
detailed estimates for a specific instrument concept in future work (Sec.~\ref{sec8}).

\subsection{Multistage Imaging: the Vernier Cascade}\label{subsec6-2}

The way around the $S \propto D^2$ scaling for the required precision is to introduce additional sets of grids, each with a different beat frequency relative to the primary (finest-pitch) grid, staggered at progressively larger beat angular scales. Each additional grid set introduces one \emph{stage} of the cascade, with progressively longer beat periods.

For an $N$-stage cascade, each stage need only achieve a smaller, \emph{per-stage disambiguation factor} $d$,
\begin{equation}
d = D^{1/N} \qquad,
\label{eq:F-per-stage}
\end{equation}%
so that combining all $N$ stages recovers the full disambiguation factor, $D = d^{N}$.
With this per-stage factor, each stage must now differentiate the alignment of only $d$ fringes rather than $D$ fringes.
Applying the same argument as Eqs.~\eqref{eq:precision-required}--\eqref{eq:S-min-D} to each stage's grid pair ($D\to d$), and following the same reasoning as
Sec.~\ref{subsec6-1} that calibration accuracy is a fixed systematic floor rather than a quantity that improves with exposure,
gives a required per-stage relative calibration accuracy satisfying the upper limit,
\begin{equation}
\epsilon \ll \frac{1}{d},
\label{eq:precision-per-stage}
\end{equation}%
and a per-stage minimum source count (statistical precision),
\begin{equation}
 S \gtrsim \left( \frac{k}{2} \right)^2 d^2 = \left( \frac{k}{2} \right)^2 D^{2/N}.
\label{eq:Ns-comparison}
\end{equation}%
The number of counts needed \emph{per stage} is much smaller than for a single-stage two-grid vernier cascade.

This $D\to d$ substitution applies uniformly at every stage, not just the first, though this requires identifying the correct fringes being compared at each stage.
At stage $m$ ($m=1,\ldots,N$), the disambiguation compares the fringes from the primary grid (grid~1) against those from grid~$m+1$,
whose beat period, by design (Sec.~\ref{subsec6-3}), satisfies $\tan(\beta_{1,m+1})=d^m\tan(\alpha_1)\approx d^m\alpha_1$.
Generalizing Eq.~\eqref{eq:beat-period} to the grid pair $(1,m+1)$: $\tan(\beta_{1,m+1})\approx\alpha_1\alpha_{m+1}/(\alpha_{m+1}-\alpha_1)$, gives $\alpha_{m+1}-\alpha_1\approx\alpha_1 d^{-m}$ for $d\gg1$ — the same small-pitch algebra as Sec.~\ref{subsec6-1}, unaffected by whether $\beta_{1,m+1}$ itself is a small or large angle.
Entering stage $m$, the surviving fringe candidates from the previous stage are spaced $d^{m-1}\alpha_1$ apart, i.e., every $d^{m-1}$th primary grid fringe
(for $m=1$ this is simply the original primary grid fringes themselves, spaced $\alpha_1$ apart, recovering Sec.~\ref{subsec6-1} exactly).
The nearest wrong fringe that still survives from the previous stage is therefore $\Delta k = d^{m-1}$ primary grid periods from the true source, and its predicted position from the primary grid
differs from grid $(m+1)$'s prediction by [cf. Eq.~\eqref{eq:fringe_spacing}],
\begin{equation}
\Delta\eta_m \approx \Delta k\,(\alpha_{m+1}-\alpha_1) \approx d^{m-1}\cdot\alpha_1 d^{-m} = \frac{\alpha_1}{d},
\label{eq:fringe-spacing-stage-m}
\end{equation}%
independent of stage $m$. The growing spacing between successive survivors exactly compensates the shrinking pitch mismatch of the later, closer-to-$\alpha_1$ grids. The precision requirement of Eq.~\eqref{eq:precision-required} therefore applies unchanged at every stage under $D\to d$, confirming that Eqs.~\eqref{eq:precision-per-stage}--\eqref{eq:Ns-comparison} hold uniformly across the cascade rather than only for its first stage. 
However, an $N$-stage cascade must get \emph{every} one of its $N$ stage decisions
right to correctly localize the source, so the confidence level of the cascade as a whole is lower than the per-stage value. We derive the
joint, whole-cascade confidence level, and its implications for the required per-stage $k$, in Appendix~\ref{appC}.

The calibration accuracy $\epsilon \sim 1/d$ of Eq.~\eqref{eq:precision-per-stage} must be met either by
manufacturing and mounting the grids to this fractional tolerance, or by an independent pre-flight or on-orbit calibration that measures the effective pitches
and separations to this precision. In particular, thermal or mechanical changes to the grid separation $L$ after calibration, e.g., from on-orbit temperature
excursions, would shift every $\alpha_j$ and $\Pi_i$ [Eqs.~\eqref{eq:alpha},\eqref{eq:phase}] together. Keeping these effects below the per-stage tolerance
$\epsilon \sim 1/d$ requires either a mechanically and thermally stable mount or a calibration procedure that tracks $L$ to the same fractional precision. We do
not model a specific structural or thermal design here, but note that this metrology and stability requirement, not only the reduced photon-counting
requirement of Eq.~\eqref{eq:Ns-comparison}, must be satisfied by any physical realization of the cascade.

We show a worked example of these numbers in Section \ref{subsec6-4}.

\subsection{Geometric Span of the Cascade}\label{subsec6-3}

The pitch of each progressive stage is arranged so that the beat periods geometrically span the full field of view. As in Eq.~\eqref{eq:beat-period}, this design condition is naturally stated in $\tan\theta$-space,
\begin{align*}
\tan(\beta_{12}) &= d\,\tan(\alpha_1) \approx d\,\alpha_1, \\
\tan(\beta_{13}) &= d^2\,\tan(\alpha_1) \approx d^2\,\alpha_1, \\
\tan(\beta_{14}) &= d^3\,\tan(\alpha_1) \approx d^3\,\alpha_1, \\
\ldots,
\end{align*}%
with the actual angular beat periods recovered as $\beta_{1j}=\arctan[\cdots]$. Because $\beta_{1j}$ grows to be comparable to $\Omega$ over the cascade, it need not itself stay small even though the underlying pitches $\alpha_j$ do (Sec.~\ref{subsec6-4} gives the worked values). The benefit of this design is that each progressive stage eliminates approximately a factor $d$ of the candidate fringes, while significantly reducing statistical requirements at each step:
\begin{align*}
\beta_{12}&: \ D \text{ candidate fringes} \ \rightarrow\ D/d \text{ viable fringes}, \\
\beta_{13}&: \ D/d \text{ candidate fringes} \ \rightarrow\ D/d^2 \text{ viable fringes}, \\
\ldots & \\
\beta_{1N+1}&: \ D/d^{N-1} \text{ candidate  fringes} \ \rightarrow\ D/d^N = 1 \text{ viable fringe.}
\end{align*}%
By cascading stages in this way, each individual stage requires substantially less accuracy and statistical precision than would be needed to resolve the full disambiguation factor $D$ in a
single stage, while the combined cascade still uniquely localizes the source to a single fringe across the entire field of view.

Beyond disambiguation, each of the $N+1$ four-phase modules in the cascade is also an independent measurement of the source position in its own right, 
so their statistics can be combined for a final localization better than any single module's own precision. If every module collects a comparable number of source
counts $S$, combining all $N+1$ measurements gives an ultimate localization precision that is close to, but never quite as good as,
\begin{equation}
\sigma_\theta \gtrsim \frac{\Delta\theta_1}{\sqrt{2(N+1)S}},
\label{eq:sigma-eta-ultimate}
\end{equation}%
the shortfall arising because the coarser-pitch modules are slightly poorer at constraining the source position than the finest grid. 
The full derivation of this positional uncertainty and its exact correction factor are given in Appendix~\ref{appB}.

\subsection{Worked Example}\label{subsec6-4}

Consider a fixed bigrid system with a FOV $\Omega = 60^\circ$, and modulation period $\alpha_1 = 1^\circ$ ($\Delta \theta = 30'$), giving a total disambiguation factor,
\begin{equation*}
D = \frac{2\tan(\Omega)}{\tan(\alpha_1)} = 198.
\label{eq:example-D}
\end{equation*}%
For this example we will assume a fixed requirement $k=3$ (per stage $k\sigma$), which results in an overall target confidence level $CL > 99\%$. (See Appendix~\ref{appC} for a more nuanced
discussion regarding requirements on $k$.)
Resolving this directly with a single-stage cascade (Sec.~\ref{subsec6-1}) would require a calibrated fringe-alignment accuracy of $\epsilon \sim 1/D =0.5\%$,
and a source count of at least $S \gtrsim \left(\frac{k}{2} \right)^2 D^2 \approx 9\times10^4$ [Eq.~\eqref{eq:S-min-D}],
which can be challenging for a photon-limited system.

Instead consider implementing this design with a three-stage cascade ($N = 3$), i.e., four grid modulation periods ($\alpha_1$ through $\alpha_4$). With three stages,
the per-stage disambiguation factor becomes,
\begin{equation*}
d = D^{1/N} \approx 198^{1/3} \approx 5.8.
\label{eq:F-example}
\end{equation*}%
With this per-stage factor, each stage must now differentiate the alignment of only $d \approx 5.8$ fringes, to a calibration accuracy of,
\begin{equation*}
\epsilon \sim \frac{1}{d},
\label{eq:precision-per-stage-example}
\end{equation*}%
This level of accuracy ($\epsilon \sim 17\%$) is readily achievable through simple calibrations.
And the number of counts needed \emph{per stage} to reach this level of precision is much smaller than for a single-stage system.
At minimum $S \gtrsim \left(\frac{k}{2} \right)^2 d^2 \approx 76$ [Eq.~\eqref{eq:Ns-comparison}], versus $S \gtrsim 9\times10^4$.

This comparison is fairest when stated in terms of the \emph{total} number of source counts collected by the instrument, since a single-stage vernier
already requires two grid pairs (primary and secondary) while the $N$-stage cascade requires $N+1$. The single-stage system therefore requires a total of
$\sim 2 \left(\frac{k}{2} \right)^2D^2 \approx 1.8\times10^5$ source counts summed over its two modules, while the three-stage cascade requires a total of only 
$\sim (N+1) \left(\frac{k}{2} \right)^2d^2 \approx 304$ summed over its four modules, a factor of $\sim600$ fewer counts even on this more conservative, total-counts basis.
A more careful analysis of these required counts on a like-for-like confidence basis (Appendix~\ref{appC}) yields a factor of
$\sim400$, comparable to the simpler comparisons above.

The pitch of each progressive stage is arranged so that the beat periods geometrically span the full field of view. Using $\tan(\beta_{1j})=d^{j-1}\tan(\alpha_1)$ and converting back to an angle via $\beta_{1j}=\arctan[\cdots]$ (Sec.~\ref{subsec6-3}),
\begin{align*}
\beta_{12} &\approx 6^\circ, \\
\beta_{13} &\approx 31^\circ, \\
\beta_{14} &\approx 74^\circ,
\end{align*}%
the last of which recovers $\arctan(2\tan\Omega)$, the true angular half-span condition of Eq.~\eqref{eq:beat-fov} applied to the full FOV. Naively taking $\beta_{14}=d^3\alpha_1\approx198^\circ$ instead, without the $\tan\theta$ correction, would overshoot a physical angle entirely. Each progressive stage eliminates approximately a factor $d \approx 5.8$ of the candidate fringes, with minimal statistical requirements at each step:
\begin{align*}
\beta_{12}&: \ 198 \text{ candidate fringes} \ \rightarrow\ 198/d \approx 34 \text{ viable fringes}, \\
\beta_{13}&: \ 34 \text{ viable fringes} \ \rightarrow\ 34/d \approx 6 \text{ viable fringes}, \\
\beta_{14}&: \ 6 \text{ viable fringes} \ \rightarrow\ 6/d \approx 1 \text{ viable fringe.}
\end{align*}%
This multi-stage vernier cascade alleviates the strict requirements on grid characterization accuracy and counting statistics, while localizing the source to
positions that are a fraction of the primary bigrid angular resolution $\Delta \theta_1$.

If each of the $N+1=4$ modules in this example collects a comparable number of counts $S$, Eq.~\eqref{eq:sigma-eta-ultimate} gives an ultimate combined
localization $\sigma_\theta \gtrsim \Delta\theta_1/(2\sqrt{2S})$. The exact combination (Appendix~\ref{appB}) shows this is achieved to within about $5\%$ for $d\approx5.8$.

\section{Discussion}\label{sec7}

\subsection{Generalization to Two Dimensions}\label{subsec7-1}

Sections~\ref{sec3}-\ref{sec6} treat localization along a single offset angle $\theta$, measured perpendicular to the grid slits. A bigrid system is insensitive to a source's position
\emph{along} its own slit direction. Two sources displaced only along that direction undergo identical transmission
(modulo the FOV envelope mentioned in Sec.~\ref{subsec4-3}), and hence identical relative count rates, in every phase channel of the cascade.
Full localization of a source on the sky therefore requires a second, independent vernier cascade whose grids are rotated $90^\circ$ relative to the first, so that its slits run parallel to the
first system's measurement axis and its own measurement axis is orthogonal to it.

Because the transmission of each system [Eq.~\eqref{eq:R-measured}] depends only on the single offset angle perpendicular to its own slits, the two orthogonal cascades are completely
independent. An $X$-cascade (slits along $\hat{y}$) recovers the source's $\theta_x$, and an identical $Y$-cascade (slits along $\hat{x}$, otherwise built to the same design)
recovers $\theta_y$, and the pair $(\theta_x, \theta_y)$ fixes the source position on the sky to within a fraction of the angular resolution $\Delta\theta_1$ of the finest stage of each
cascade [Eq.~\eqref{eq:sigma-eta-ultimate}]. No joint two-dimensional fit or deconvolution is required. Because the two measurement axes share no grids, detectors, or fringe pattern,
the two-dimensional localization problem factors exactly into two independent copies of the one-dimensional problem solved in Secs.~\ref{sec5}--\ref{sec6},
and the resulting position uncertainty is simply the quadrature combination of the two 1D uncertainties. A complete two-dimensional, $N$-stage, four-phase instrument therefore
comprises two orthogonal cascades, each with $N$ stages and four bigrid/detector channels per stage, for $8(N+1)$ simple photon-counting channels in total,
e.g., 32 channels for the $N=3$ worked example of Sec.~\ref{subsec6-4}. A third orientation (e.g., at $45^\circ$ to the first two) can optionally be added as a cross-check
on the aspect solution and as additional redundancy (Sec.~\ref{subsec7-2}), following the precedent of earlier multi-orientation RMC designs,
but is not required for a unique two-dimensional localization.

\subsection{Advantages of a Fixed, Multi-Phase Vernier-Cascade Architecture}\label{subsec7-2}

Relative to rotating modulation collimators such as WATCH, whose wide field of view is achieved by scanning a single bigrid pair in time, the
architecture developed here retains the wide field of view that motivated the WATCH RMC instruments while extending it, via the vernier cascade,
to finer source localization without requiring rotation. Because the field of view is set purely by the grid geometry [Eq.~\eqref{eq:omega}]
rather than by a rotation duty cycle or scanning schedule, the instrument is wide-field \emph{by construction}: every source within $\pm\Omega$ of boresight
in both axes is modulated, and hence localizable, at every instant, with no limitations due to rotational sampling timescales.

The architecture is also highly redundant. A full instrument comprises $8(N+1)$ (or $12(N+1)$, with a third orientation) independent, simultaneously-read-out channels,
and both the four-phase demodulation of Sec.~\ref{subsec5-3} and the multistage cascade of Sec.~\ref{sec6} are themselves forms of redundancy. The four-phase
combination [Eq.~\eqref{eq:four-phase-ratio}] already double-differences four channels to cancel background and leakage systematics, and the
longer-beat-period stages of the cascade provide an independent (if lower-precision) cross-check on the position implied by finer stages. The loss or degradation of a single bigrid or detector need not be
catastrophic: it degrades the statistical precision or coverage of one stage (reverting to two-phase or single-phase demodulation) rather than disabling the instrument.

Coded-aperture imagers such as \emph{Swift}/BAT, HETE-2/WXM, or SVOM/ECLAIRs also require no rotation. 
However, since the source phase is derived purely from ratios and differences of simultaneous, scalar count rates [Eq.~\eqref{eq:four-phase-ratio}]
rather than from a spatially resolved image, each channel requires only a simple, non-imaging photon-counting detector such as a scintillator with photomultiplier or silicon
photomultiplier readout. This is a substantially lower-mass, lower-power, and lower-complexity detector requirement than the highly pixelated, position-sensitive detector
arrays needed by these coded-aperture imagers \citep{Gehrels2004,Kawai2003,Godet2014}, and the vernier cascade further requires far fewer, and far less
stringently toleranced, resolution elements than the fine, high-contrast mask patterns these instruments depend on.

Relative to non-imaging, count-rate localizing monitors such as BATSE, Fermi-GBM, and BurstCube, arguably
the closest cost and complexity to the architecture developed here (Sec.~\ref{sec2}), such instruments also use simple, non-imaging photon
counters and require no rotation, but recover only a coarse direction from the relative count rates among several unaimed detectors, typically to a few
degrees. The architecture developed here instead uses collimating grids to encode angular position directly in each channel's modulation. For a comparable detector area,
the fixed modulation collimator design reaches substantially finer localization (Sec.~\ref{subsec6-4}) at the cost of reduced detection sensitivity due to the average grid
transmission ($25\%$) and the additional complication of
grid hardware and calibration these monitors do not require.

Relative to precision RMCs such as RHESSI, the vernier cascade also relaxes the demands placed on grid fabrication relative to a single-stage cascade or previous RMC instruments.
In the architecture developed here, no grid pitch need be manufactured
to better than the accuracy with which it is subsequently calibrated (Sec.~\ref{subsec6-1}), and the multistage cascade construction (Sec.~\ref{subsec6-2})
shows that this required accuracy relaxes rapidly with the number of stages $N$.
For context, RHESSI's finest subcollimators have a
grid pitch of 34~\textmu m \citep{Hurford2002}, which is significantly finer than the grid pitches required for GRB localization.
Furthermore, because the grids are fixed rather than spinning,
this accuracy also need not be maintained against decades of bearing wear, dynamic imbalance, or spin-axis drift.

Again relative to RMCs, because every channel is read out simultaneously and continuously, the architecture requires no time modulation to build up a position measurement.
An RMC necessarily integrates over some fraction of a rotation period to accumulate a usable modulation signal (e.g., WATCH's $\sim$1~Hz spin corresponds
to a modulation period of $\sim$1~s). A source that varies significantly within a
rotation period, as GRBs routinely do on timescales down to milliseconds, has its modulation pattern smeared and its reconstructed phase (and hence position) biased
by that variability. The four-phase measurement of Eq.~\eqref{eq:four-phase-ratio} requires only a single set of simultaneous count-rate samples across four
co-pointed channels, with no assumption about the source variability. The architecture is therefore intrinsically
well suited to impulsive, rapidly varying transients.

\subsection{Persistent Sources and Survey Science}\label{subsec7-25}

As presented in this paper, demodulation of a transient GRB signal would be challenged by persistent bright sources within the FOV.
A coded-mask imager forms an image of its field of view, so persistent bright sources present in that field are naturally
separated from a transient by their distinct, unchanging position in the reconstructed image. For the fixed modulation collimator architecture developed here, 
a persistent source in the field of view produces its own steady modulated count rate in every channel, which does not cancel in the
four-phase differences [Eq.~\eqref{eq:four-phase-ratio-counts}] the way a constant, unmodulated background does, and so must
instead be removed by either pre-burst rate subtraction or requiring the persistent source population to be tracked.
One way to track these persistent sources is to slowly raster the pointing axis of the fixed modulation collimator instrument, 
which would modulate the persistent source rates over time, enabling localization and flux measurement of these sources.
This would also enable survey science in addition to transient localization. 
Such a raster occurs naturally for a zenith-pointed spacecraft in low Earth orbit, for example.
Technically, two simultaneous transients within the field of view could be difficult to
disentangle from a single set of phase measurements, since the demodulation of Sec.~\ref{sec5} assumes a single dominant modulated source per cascade.
However the likelihood for multiple GRBs to occur simultaneously in the field of the instrument is vanishingly small.

\subsection{A Candidate for Time-Domain and Multi-Messenger Astrophysics}\label{subsec7-3}

Taken together, wide field of view, redundancy, simple non-imaging detectors, modest grid-fabrication requirements, and localization on a timescale
set by the source rather than by instrument rotation make the fixed vernier-cascade modulation collimator a strong candidate architecture for
time-domain and multi-messenger astrophysics (TDAMM) \citep{NASEM2021}. A central observational challenge for TDAMM is the rapid identification
of high-energy electromagnetic counterparts to gravitational-wave and neutrino events, most of which are localized by their discovery facilities only to regions
of many to hundreds of square degrees. A wide-field, always-on, self-localizing gamma-ray instrument that requires no rotating mechanism,
no position-sensitive imaging detector, and no extended integration to build up a position measurement is well matched to a small, low-cost satellite,
or a constellation of them, dedicated to rapid electromagnetic counterpart searches to anchor multi-messenger follow-up observations.

\section{Conclusion}\label{sec8}

We have developed a fixed, non-rotating modulation-collimator architecture for wide-field localization of gamma-ray bursts and other rapidly varying transients.
Building on the two-grid vernier scale of \citet{Gursky1966}, we showed that a cascade of $N$ bigrid stages with progressively longer beat periods, each read out through
four-phase demodulation, converts a purely local fringe measurement [Eq.~\eqref{eq:four-phase-ratio}] into an unambiguous source position across an arbitrarily
wide field of view $2\Omega$, while relaxing the per-stage calibration accuracy and photon-counting requirements from $S \propto D^2$ for a single vernier pair
to $S \propto D^{2/N}$ for an $N$-stage cascade. In the worked example of Sec.~\ref{subsec6-4}, a $\pm60^\circ$ field of view with $30'$ angular resolution,
which would require calibration accuracy $\sim 0.5\%$ and $S \gtrsim 9 \times 10^4$ source counts to disambiguate with a single-stage system, is instead resolved with three
cascaded stages, each requiring only $\sim$17\% accuracy and $\sim$76 source counts. Generalized to two dimensions with a pair of orthogonal cascades
(Sec.~\ref{subsec7-1}), the same $N=3$ example yields wide-field localization with $8(N+1) = 32$ simple, non-imaging photon-counting channels, no moving parts,
and no time modulation, making it well suited to impulsive, rapidly varying sources on 3-axis-stabilized platforms (Sec.~\ref{sec7}).

The treatment developed here is intentionally introductory, and several simplifications should be kept in mind when assessing these results. The transmission
function of Sec.~\ref{sec4} assumes an idealized triangular (sawtooth) fringe from perfectly parallel, uniformly pitched slits; real grids will exhibit
manufacturing tolerances, slit-edge diffraction, and thermal or mechanical distortions that broaden or otherwise modify this fringe shape, and the achievable
demodulation precision should ultimately be validated against a full grid-response model rather than the simple transmission function used here. The performance parameters
of Eqs.~\eqref{eq:dtheta}--\eqref{eq:omega} and the transmission function of Sec.~\ref{sec4} were also evaluated at a single, representative photon energy;
a real grid's opacity, and hence its modulation amplitude $a$ and leakage term $\ell$, are energy-dependent, so the geometric design developed here must
be paired with an energy-resolved sensitivity analysis before it can be translated into hardware requirements for a specific energy band. Finally, we have not
addressed the detector background and its energy dependence in any detail, nor the systematic effects of finite detector size and finite grid-pair
alignment tolerances, all of which will set the practical floor on the achievable localization precision.

These simplifications point directly to the natural next steps for this work. A laboratory demonstration of a single bigrid pair, and ultimately of a multistage
cascade, would validate the demodulation and disambiguation methods of Secs.~\ref{sec5}--\ref{sec6} against real grid hardware and real background conditions.
A companion manufacturing tolerance study, informed by the relaxed per-stage accuracy requirements of Sec.~\ref{subsec6-1}, would establish whether the
grid pitches required for GRB-scale localization can be met with low-cost fabrication techniques such as 3D-printed tungsten of standard photochemical etching. A full instrument-response
simulation, incorporating a realistic energy-dependent grid response, detector background, and orbital or attitude environment, would translate the geometric
disambiguation factor $D$ developed here into a concrete sensitivity and localization-accuracy predictions for a specific mission concept. We regard the fixed
vernier cascade architecture developed in this paper as a promising and low-complexity candidate for a dedicated, wide-field, continuous burst localization
instrument, and we intend to pursue these validation steps toward that end. We have prepared a companion paper addressing optimization of the grid geometry for wide-field 
applications which studies self-vignetting effects introduced in Sec.~\ref{subsec4-3} in detail \citep{Boggs2026},
which is beyond the scope of the basic technique presented here.
We are pursuing a focused follow-up study 
to define a specific strawman design and carry it through the effective area, background, and sensitivity analysis called for above. We will report the detailed
on-sky performance estimates of this strawman design in a subsequent paper.

\appendix

\section{Statistical Precision of Four-Phase Demodulation}\label{appA}

In Section~\ref{subsec5-3} we demonstrated that a single four-phase demodulation gives a clean, background- and leakage-insensitive measurement of the source phase $\phi$, with
statistical precision $\sigma_\eta=\Delta\theta/(\sqrt{2S})$ [Eq.~\eqref{eq:sigma-eta-four}] on the fringe locations, where $S$ is the total number of source counts collected by the single four-phase
system of angular resolution $\Delta\theta$.
Here we derive that result. We work under the simplifying assumption of adopting the ideal (perfectly opaque grid) transmission model [Eq.~\eqref{eq:trans-ideal}], rather than the more general leakage model of Eq.~\eqref{eq:trans-real}. Under this assumption, the four channel counts are pure Poisson measurements of the transmitted source flux plus the background.

For an exposure time $\tau$, the expected number of source counts in channel $i$ is
\begin{equation}
S_i = T(\phi - \Pi_i)\,\mathcal{F} A\,\tau, \qquad i = 1,2,3,4,
\label{eq:Si-counts}
\end{equation}%
with $\Pi_1=0^\circ$, $\Pi_2=90^\circ$, $\Pi_3=180^\circ$, $\Pi_4=270^\circ$ as in Sec.~\ref{subsec5-3}, and each $S_i$ an independent Poisson random variable with $\mathrm{Var}(S_i) = S_i$.
First, note a particularly useful feature of this combination.
The symmetry of the ideal transmission function is the exact identity,
\begin{equation}
T(\phi) + T(\phi - 180^\circ) = \frac{1}{2} \qquad \text{for all } \phi,
\label{eq:trans-identity}
\end{equation}%
which follows by direct substitution into Eq.~\eqref{eq:T-ideal} using the symmetry of $h(\phi)$.
Applying Eq.~\eqref{eq:trans-identity} to the opposing channel pairs $(1,3)$ and $(2,4)$ gives the total count rate collected across all four channels,
\begin{equation}
R \equiv R_1+R_3+R_2+R_4 = \mathcal{F}A\Big[\big(T(\phi){+}T(\phi{-}180^\circ)\big) + \big(T(\phi{-}90^\circ){+}T(\phi{-}270^\circ)\big)\Big] = \mathcal{F}A,
\label{eq:R-total}
\end{equation}%
independent of the source phase $\phi$. So the combined count rate of the four collimated detectors in a four-phase module is equal to the count rate
we would expect from a single uncollimated detector of area $A$. Integrated over the exposure time $\tau$, this leads to the relation $S_1+S_3 = S/2$ and $S_2+S_4 = S/2$, where $S = \mathcal{F}A\tau$.

We can also define the total number of background counts in each channel, $B_i = \mathcal{B}_i \tau$, each an independent Poisson random variable with $\mathrm{Var}(B_i) = B_i$, and the total background counts for the four-phase system,
\begin{equation}
B = B_1 + B_2 + B_3 +B_4
\label{eq:B-total}
\end{equation}%
where in the derivation below we will assume $B_i = B/4$.

We now define our four-phase demodulation factor $\Gamma$ [Eq.~\eqref{eq:four-phase-ratio}] in terms of channel counts instead of rates,
\begin{equation}
\Gamma = \frac{(S_1 + B_1) - (S_3 + B_3)}{(S_2 + B_2) - (S_4 + B_4)} \equiv \frac{D_{13}}{D_{24}}.
\label{eq:four-phase-ratio-counts}
\end{equation}%
where we have defined the channel-pair count differences $D_{13} \equiv (S_1 + B_1) - (S_3 + B_3)$ and $D_{24}\equiv (S_2 + B_2) - (S_4 + B_4)$. Because all $S_i$ and $B_i$
are mutually independent, $\mathrm{Var}(D_{13}) = S_1+S_3+B_1+B_3$ and $\mathrm{Var}(D_{24}) = S_2+S_4+B_2+B_4$. Combining these pairs gives,
\begin{equation}
\mathrm{Var}(D_{13}) = \mathrm{Var}(D_{24}) = \frac{(S+B)}{2},
\label{eq:var-D}
\end{equation}%
again exactly, and again independent of $\phi$.

The four-phase demodulation ratio [Eq.~\eqref{eq:four-phase-ratio}] is the ratio of these two independent, equal-variance quantities, $\Gamma \equiv D_{13}/D_{24} = h(\phi)/h(\phi-90^\circ)$. Standard propagation of errors for the ratio of two independent random variables gives
\begin{equation}
\sigma_\Gamma^2 = \frac{\mathrm{Var}(D_{13})}{D_{24}^2} + \frac{D_{13}^2\,\mathrm{Var}(D_{24})}{D_{24}^4} = \frac{(S+B)}{2}\,\frac{1+\Gamma^2}{D_{24}^2}.
\label{eq:sigma-Q}
\end{equation}%

We evaluate Eq.~\eqref{eq:sigma-Q} at the representative phase $\phi=45^\circ$ ($=\pi/4$), exactly midway between the phase origins of channels 1 and 2,
where the two channel-pair differences carry comparable statistical weight. From Eq.~\eqref{eq:trans-identity},
\begin{align}
D_{13} &= S\left[2T(\phi)-\tfrac12\right], &
D_{24} &= S\left[2T(\phi-90^\circ)-\tfrac12\right],
\label{eq:D13D24-T}
\end{align}%
and $T(45^\circ)=T(-45^\circ)=3/8$ [Eq.~\eqref{eq:trans-ideal}], giving $D_{13}=D_{24}=S/4$ and $\Gamma=1$. Differentiating Eq.~\eqref{eq:D13D24-T} via the quotient rule, using $dT/d\phi = \mp 1/2\pi$ for $\phi \gtrless 0$ [Eq.~\eqref{eq:trans-ideal}], gives $|d\Gamma/d\phi| = 8/\pi$ at this phase, so that
\begin{equation*}
\sigma_\Gamma = \sqrt{\frac{(S+B)}{2}\cdot\frac{1+1^2}{(S/4)^2}} = \frac{4\sqrt{(S+B)}}{S}, \qquad
\sigma_\phi = \frac{\sigma_\Gamma}{|d\Gamma/d\phi|} = \frac{\pi \sqrt{(S+B)}}{2S}.
\end{equation*}%

This result is specific to the choice $\phi=45^\circ$. Repeating the calculation of Eq.~\eqref{eq:sigma-Q} and $|d\Gamma/d\phi|$ at every phase shows that $\sigma_\phi(\phi)$ varies smoothly and periodically, with period $90^\circ$ (set by the channel spacing): it reaches the value derived above at every $\phi=45^\circ+90^\circ n$ and rises to a maximum exactly $\sqrt2$ times larger at the channel phase origins $\phi=90^\circ n$, where one channel-pair difference vanishes. This factor of $\sqrt2$ is independent of $S$ and $B$. Adopting the maximum as a conservative, phase-independent bound gives
\begin{equation}
\sigma_\phi \le \sqrt{2}\,\frac{\pi \sqrt{(S+B)}}{2S}.
\label{eq:sigma-phi}
\end{equation}%

Converting back to the fringe offset via Eq.~\eqref{eq:phase}, $\sigma_\delta = (\alpha/2\pi)\,\sigma_\phi$, and using $\Delta\theta = \alpha/2$ [Eq.~\eqref{eq:dtheta}], the uncertainty on the recovered fringe position $\eta_k$ [Eq.~\eqref{eq:fringes-single}] is bounded by
\begin{equation}
\sigma_\eta = \sigma_\delta \le \sqrt{2}\,\frac{\Delta\theta\sqrt{(S+B)}}{2S}.
\label{eq:sigma-eta-four-app-general}
\end{equation}%
In the source-dominated limit $(S \gg B)$, this bound simplifies to,
\begin{equation}
\sigma_\eta = \sigma_\delta \le \frac{\Delta\theta}{\sqrt{2S}},
\label{eq:sigma-eta-four-app}
\end{equation}%
which we adopt as the conservative bound for $\sigma_\eta$ used in Eq.~\eqref{eq:sigma-eta-four} in the main text.

\section{Combined Cascade Localization}\label{appB}

Section~\ref{subsec6-3} states that combining the position measurements of all $N+1$ four-phase modules in a cascade, each with $S$ source counts, gives an ultimate localization precision close to, but slightly worse than, $\Delta\theta_1/(\sqrt{2(N+1)S})$ [Eq.~\eqref{eq:sigma-eta-ultimate}]. Here we derive that result and its exact and approximate correction factor.

Once the vernier cascade has resolved the fringe ambiguity, every one of the $N+1$ modules (fringe periods $\alpha_j$) provides an independent, 
unbiased measurement of the same true source position $\theta$, each with its own four-phase statistical precision [Eq.~\eqref{eq:sigma-eta-four}],
\begin{equation}
\sigma_{\eta,j} = \frac{\alpha_j}{2\sqrt{2S_j}}, \qquad j=1,\ldots,N+1,
\label{eq:sigma-eta-j}
\end{equation}%
where $S_j$ is the number of source counts collected by module $j$, and we will assume $S_j = S$ for each four-phase module.
Because the $N+1$ measurements are statistically independent, the standard inverse-variance combination of independent unbiased estimators of a common parameter applies
when determining the net uncertainty on the source location, $\sigma_{\theta}$,
\begin{equation}
\frac{1}{\sigma_{\theta}^2} = \sum_{j=1}^{N+1} \frac{1}{\sigma_{\eta,j}^2} = \frac{8 S}{\alpha_1^2}\sum_{j=1}^{N+1} \left(\frac{\alpha_1}{\alpha_j}\right)^2,
\label{eq:Seff-def0}
\end{equation}%
where for the multi-stage vernier cascade we have defined,
\begin{equation}
\frac{\alpha_1}{\alpha_j} = 1-d^{-(j-1)}, \qquad j=2,\ldots,N+1,
\label{eq:alpha-ratio}
\end{equation}%
(with the trivial $j=1$ case, $\alpha_1/\alpha_1=1$, handled separately below). Note that every module's pitch lies close to $\alpha_1$, 
approaching $\alpha_1$ as $j$ increases and the beat period with the primary grid grows.
Plugging this relation for $\alpha_1/\alpha_j$ back into [Eq.~\eqref{eq:Seff-def0}]:
\begin{equation}
\frac{1}{\sigma_{\theta}^2} =  \frac{8 S}{\alpha_1^2}\left[1+\sum_{m=1}^{N}\left(1-d^{-m}\right)^2\right].
\label{eq:Seff-def1}
\end{equation}%
Using the finite geometric-series sum $\sum_{m=1}^{N}d^{-m}=(1-d^{-N})/(d-1)$, we can write this relation as,
\begin{equation}
\frac{1}{\sigma_{\theta}^2} =  \frac{8 S}{\alpha_1^2}\left[(N+1) - \frac{2\left(1-d^{-N}\right)}{d-1} + \frac{1-d^{-2N}}{d^2-1}\right].
\label{eq:Seff-def2}
\end{equation}%
Assuming $d\gg1$ and replacing $\Delta \theta_1 = \alpha_1/2$, this reduces to the simple
approximation,
\begin{equation}
\sigma_{\theta} \approx \frac{\Delta\theta_1} {\sqrt{2(N+1 -\frac{2}{d-1})S}},
\label{eq:Seff-approx}
\end{equation}%
with the correction effectively dominated by the $\alpha_2$ module, the module with the coarsest angular resolution.

Applying Eq.~\eqref{eq:Seff-def2} to the worked example of Sec.~\ref{subsec6-4} ($N=3$, $d\approx5.8$) we find,
\begin{equation}
\sigma_{\theta} \approx 1.05 \times \frac{\Delta\theta_1}{\sqrt{2(N+1)S}},
\end{equation}%
i.e., the achievable combined localization is about $5\%$ worse than the idealized bound of Eq.~\eqref{eq:sigma-eta-ultimate}, with the shortfall set almost entirely by the second module's
relatively degraded angular resolution with respect to the primary grid.

\section{Joint Disambiguation Confidence Across a Multistage Cascade}\label{appC}

Section~\ref{subsec6-1} introduced the disambiguation criterion (i.e., misalignment, Eq.~\eqref{eq:S-min-D}) for a single neighboring fringe
at the $k \sigma$ level, giving $S \gtrsim (kD/2)^2$ for a single
vernier pair and $S \gtrsim (kd/2)^2$ per stage of an $N$-stage cascade. That result describes a single pairwise decision, at a single stage, against a single
neighboring fringe. Because the cascade only correctly localizes the source if \emph{every} one of its $N$ stages discriminates the correct fringe against
\emph{both} of its neighbor fringes, the confidence level of the cascade as a whole is generally lower than the per-stage value. We can derive a simple estimate
of this whole-cascade confidence level ($CL$).

For a single neighboring fringe measured at the $k \sigma$ level, the probability ($f$) that this fringe is \emph{misidentified} as aligned is given by,
\begin{equation}
f = 1-\Phi(k),
\label{eq:single-fail}
\end{equation}%
Where $\Phi(x)$ is the cumulative distribution function (CDF) of the standard normal distribution:
\begin{equation}
\Phi(x)= \frac{1}{\sqrt{2\pi}} \int_{-\infty}^{x} e^{-t^2/2}\,dt.
\label{eq:CDF}
\end{equation}%
This is the one-sided CDF because the neighboring fringe will only be misidentified as aligned if the statistical fluctuations make the separation between the fringes smaller, not larger. 
Given this individual failure rate $f$, we could introduce the confidence level that this individual neighbor is correctly identified as misaligned as $CL = 1-f$. 
However, we are really interested in \emph{all} the neighbor fringes. Each of the $N$ stages has two such neighbors, one on either side of the correct fringe, so there
are $2N$ distinct ways the cascade can misidentify the source fringe at some stage. Assuming these $2N$ possible failures are uncorrelated and that 
the individual failure rate is small ($f \ll 1$), the combined probability that \emph{one} of the $2N$ neighbor fringes is misidentified as aligned is $2Nf$,
which we can use to derive an an overall confidence level that the correct fringe has been disambiguated across the entire multistage cascade:
\begin{equation}
CL = 1-2Nf = 1-2N [1-\Phi(k)].
\label{eq:CL}
\end{equation}%

For an overall target confidence level, Eq.~\eqref{eq:CL} provides an implicit function that can be solved numerically to determine the corresponding value of $k$. 
An important feature of this equation is that $k$ scales relatively weakly with $N$. 
Table~\ref{tab:cascade-confidence} evaluates Eq.~\eqref{eq:CL} for different values of the target $CL$ for 
the $N=3$ worked example of Sec.~\ref{subsec6-4} compared to a single-stage vernier $N=1$.

\begin{table*}[h]
\centering
\footnotesize
\begin{tabular}{@{}cccccc@{}}
\toprule
Target CL & $k$ (N=3) & $S$/stage & $S$ total (N=3) & $k$ (N=1) & $S$ total (N=1) \\
\midrule
$68.3\%$ & $1.62$ & $22$  & $89$ & $1.00$ & $1.97\times10^4$ \\
$95\%$   & $2.40$ & $49$ & $196$ & $1.96$ & $7.56\times10^4$ \\
$99\%$   & $2.94$ & $73$ & $294$ & $2.58$ & $1.23\times10^5$ \\
\bottomrule
\end{tabular}
\caption{Per-stage $k$ required for a target whole-cascade confidence level for $N=3$ ($d\approx5.8$, $D\approx198$).
The corresponding counts $S$ per stage and total are given as well. 
The last two columns give these parameters for a single-stage ($N=1$), two-grid vernier reaching the same target confidence level.}
\label{tab:cascade-confidence}
\end{table*}

Requiring a fixed whole-cascade confidence level appropriately raises the required per-stage $k$, and hence $S$, above the flat,
per-decision value used in Sec.~\ref{subsec6-4} ($k=3$, $S \gtrsim 304$). As shown in Table~\ref{tab:cascade-confidence},
reaching $99\%$ confidence that the entire cascade locks onto the correct
source fringe requires $S\gtrsim294$ total. Evaluating the single-stage, two-grid vernier at this same $99\%$ confidence (i.e., $N=1$ with $D$
in place of $d$) requires $S\gtrsim1.23\times10^5$ total, so the cascade's advantage, now compared on a like-for-like confidence basis, remains a factor of $\sim 400$.

\begin{acknowledgments}
During the preparation of this manuscript, the author used Claude (Anthropic, Claude Sonnet 5) to assist with editing text, checking internal consistency, and verifying statistical derivations in Secs. 5–6 and Appendix A. The ultimate responsibility for the scientific content and integrity of this work rests solely with the human author. Thanks to E. Burns and C. Kierans for helpful feedback and context,
and to an anonymous referee for constructive insights and suggestions for improvement. 
\end{acknowledgments}

\bibliography{refs}
\bibliographystyle{aasjournalv7.1}

\end{document}